\documentstyle[12pt]{article}
\begin{document}
\title{The Crumpling Transition of Dynamically Triangulated Random Surfaces}

\author{ \\ {} \\  Christian M\"unkel and Dieter W. Heermann\\ {} \\ }
 \date
     {{\em
        Institut f\"ur theoretische Physik \\
        Universit\"at Heidelberg \\
	Philosophenweg 19\\
	6900 Heidelberg\\
	and\\
	Interdisziplin\"ares Zentrum\\
	f\"ur wissenschaftliches Rechnen\\
	der Universit\"at Heidelberg\\
        Germany
     }}
  \vspace {3ex}

\maketitle

\begin{abstract}
We present the crumpling transition in three-dimensional Euclidian space
of dynamically triangulated random surfaces with edge extrinsic curvature
and fixed topology of a sphere
as well as simulations of a dynamically triangulated torus.
We used longer runs than previous simulations and give new and
more accurate estimates of critical exponents.
Our data indicate a cusp singularity in the specific heat. The transition
temperature, as well as the exponents are topology dependent.
\end{abstract}

\vspace{1cm}
\vfill\eject

\section{Introduction}
\par In this paper we are concerned with the statistical mechanics of
surfaces.
A possible approach in
space dimensions $D>1$ is to discretize the surface using a triangulation
[1-4].
For this we need to specify the Hamiltonian ${\cal H}$.
The surface $S$ is replaced by a simplical triangulation $T$,
specified by the number of nodes $N$, of links and triangles, and the
$X$-coordinate field by the coordinates $X$ of the nodes. The metrical
fluctuations of the manifold are modeled by summing over triangulations
induced by link-flips
[5-7].
The
Hamiltonian is now choosen, such that the partition function is not dominated
by configurations with spikes. In order to surpress these spikes one
adds a term with extrinsic curvature.
As a function of the extrinsic curvature
the model shows a transition
(crumpling transition) at finite rigidity of the surface
[8-18].
In a previous paper, we discussed the
extrinsic and intrinsic geometrical properties of dynamically triangulated
random surfaces
(for example the Hausdorff-dimension, spectral and spreading dimension)
above, below and at the transition \cite{muenkel92a}.

The partition function
of the above model can be written as

\begin{equation}
{\cal Z}_N = \int{d^DX_0 {\int {\prod_{i = 1}^{N-1}{d^DX_i e^{-{\cal H}}}}}}
\end{equation}

\noindent where the translational mode is integrated out.
The Hamiltonian ${\cal H}$ is defined as

\begin{equation}
{\cal H} = \ \beta \cdot \underbrace{\sum_{\langle i,j \rangle}^{N}{(X_i^\mu -
X_j^\mu)^2}}_{{\cal H}_g}
  + \ \lambda \cdot \underbrace{\sum_{\bigtriangleup_i,\bigtriangleup_j}{(1 -
      \hat{n}_{\bigtriangleup_i} \cdot \hat{n}_{\bigtriangleup_j})}}_{{\cal
H}_e}
  - \ \alpha \cdot \underbrace{\sum_{i = 0}^{N}{\log{\sigma_i}}}_{{\cal H}_m} \
\;.
\label{equ:Hamiltonian_dtrs}
\end{equation}

\noindent The Gaussian part of the Hamiltonian ${\cal H}_g$
is a sum over the positions $X$ in embedding Euclidian space of all nearest
neighbours nodes, i.e. all links of the triangulation.
We shall use $\beta=1$ because of the rescaling invariance.

\noindent ${\cal H}_e$ is an edge extrinsic curvature term
[19-26].
$\sum_{\bigtriangleup_i,\bigtriangleup_j}$ denotes a summation over all
adjacent triangles which share an edge and
$\hat{n}_{\bigtriangleup_i} \cdot \hat{n}_{\bigtriangleup_j}$ is the scalar
product of the vectors normal to a triangle.

\noindent The third part of the Hamiltonian ${\cal H}_m$ is
the discretization of the square root of the metric.
$\sigma_i$ denotes the number of nearest
neighbours of node $i$. $\alpha$ depends on the measure. We used
$\alpha=D/2$ for the simulations in $D=3$ dimensional embedding space.

Here we investigate such a model numerically (Monte Carlo simulations),
applying finite-size
scaling ideas.  These ideas have proven successful in other
contexts for the analysis of phase transitions. We expect
that the above model shows a phase transition at some critical
value of the coupling $\lambda$. There are several  questions
which we want to address: What is the order of the transition? If the
transition is of second order, what are the exponents? Are the exponents
topology dependent? Is the transition temperature topology dependent?

\section{Simulation Method and Autocorrelations}

First let us study the
effect of the extrinsic curvature on the auto-correlation time of
several observables, i.e., we are interested in the dynamics of a random
surface induced by a Monte Carlo process.
Such a study is of course a prerequisite for a detailed numerical study
of the apparent transition which such surfaces undergo as a function of
a bending rigidity. From our results which we present below we must
conclude that extensive computer resources must be applied because of
the very long relaxation times which even for moderate surface sizes
can reach several tenthousand sweeps.

We want to look at the closed dynamically triangulated random surfaces without
self-avoidance.
As models
for such surfaces we take the first two topologically closed surfaces:
The sphere and the torus (c.f. Figure~\ref{fig:show}).
The torus is specified by identifying edges
of the parameter space ${\cal P}$ whereas
the sphere, because of its genus needs a slightly more complicated
set-up procedure.

\begin{figure*}[htbp]
\caption{\em Shown are two examples of configurations of dynamically
             triangulated random surfaces. The left picture shows a
             sphere and the right part a torus.
         }
\label{fig:show}
\end{figure*}

What we are interested in is to calculate the auto-correlation function
of an observable ${\cal O}$ as a function of time
\begin{equation}
 \rho_{\cal O }(t,N) = \frac{<{\cal O}(0){\cal O}(t)>_N - <{\cal O}>_N^2}
                           {<{\cal O}^2>_N - <{\cal O}>_N^2}
\end{equation}
\noindent and its dependence on the number of nodes $N$ of the surface.
{}From the
auto-correlation function we are then able to extract the auto-correlation
time $\tau_{\cal O} (N)$. This is done calculating
the integrated auto-correlation function, i.e.,
\begin{equation}
\tau_{int,{\cal O}} = \frac{1}{2} \int_{-\infty}^{+\infty} dt\ \ \rho_{\cal
O}(t,N) \ \ \ .
\end{equation}
We also compared the integrated auto-correlation time with the exponential
correlation time and the statistical inefficiency, but found no disagreement
within the errors.
If the successively generated
configurations can be considered as non-interacting beads as is done
in the Rouse theory
\cite{rouse53,edwards86} we will find for example for the radius of gyration
\begin{equation}
\tau_{{\cal R}_{gyr}^2} \propto N
\end{equation}
and in general we expect
\begin{equation}
\tau_{\cal O} \propto N^a \ \ \ .
\end{equation}

\par To calculate observables such as the gaussian part ${\cal H}_g$ of the
Hamiltonian
or
the radius of gyration $R_{gyr}$,
we use the standard Monte Carlo algorithm
[29-31].
Such an algorithm induces a stochastic dynamics and all our results apply
only to such an induced dynamics.
One Monte Carlo sweep is completed when each triangulation point was given
the chance for a displacement from its previous position and the edges
of the triangulation were given the chance to re-connected or flip
to an orthogonal position (c.f. Figure~\ref{fig:mc}.
The flip operation implements the sum over all
possible triangulation). This corresponds
to one new configuration.

\begin{figure*}[htbp]
\caption{\em This figure demonstrates the elementary moves which are made
             during one update of a
             triangulated random surfaces.
         }
\label{fig:mc}
\end{figure*}

Table~\ref{tab_sph} summarizes our results for the
exponent $a$ for the sphere.
The exponents where extracted from the
data of system sizes from $36$ up to $288$.
At this point we should mention that the expected transition is at $\lambda_c
\approx
1.5$. Out data was collected above and below the transition.

Above and below the transition the radius of gyration shows the Rouse
behaviour. The gaussian part of the
Hamiltonian shows clearly a dependence on the bending rigidity. Above the
transition point the auto-correlation exponent changes to one and below
it is almost zero. The same holds for the edge part of the Hamiltonian.

\begin{table}[htbp]
\center
\begin{tabular} {|c||c|c|c||c|c|}  \hline
$\lambda$ & \multicolumn{3}{|c||}{correlation time $\tau$ of ...} &
\multicolumn{2}{|c|}{acceptance rate} \\ \cline{2-6}
 & $\tau_{{\cal H}_{gaussian}}$ & $\tau_{{\cal H}_{edge}}$ & $R_{gyr}^2$ &
shift & flips \\  \hline \hline
0.00 & $0.2 \pm 0.3$ & $0.2 \pm 0.2$ & $1.1 \pm 0.2$ & 0.51 & 0.76 \\ \hline
1.00 & $0.1 \pm 0.3$ & $0.4 \pm 0.2$ & $1.1 \pm 0.4$ & 0.51 & 0.36 \\ \hline
1.75 & $1.0 \pm 0.1$ & $0.8 \pm 0.2$ & $1.1 \pm 0.2$ & 0.47 & 0.28--0.34
\\ \hline
3.00 & $1.0 \pm 0.1$ & $1.0 \pm 0.1$ & $1.0 \pm 0.3$ & 0.50 & 0.23--0.33
\\ \hline
\end{tabular}
\caption{\em Table of the results for the exponent $a$ on the auto-correlation
times of dynamically
             triangulated random sphere. $\lambda$ is the measure of the
             bending rigidity. \mbox{\em Shift} and \mbox{\em flips} give the
             acceptance probability
             for a displacement of a node and flip for an edge change.
         }
\label{tab_sph}
\end{table}

Almost all of the  observations for the spherical case carry over
to the case of the torus.

\section{Results of the Simulations}

\noindent Finite size scaling assumes, that there is only one relevant linear
length scale, which is compared to the correlation length. To apply finite
size scaling to the crumpling transition of dynamically triangulated
random surfaces (DTRS) one must therefore assume
a single length scale determined by the number of nodes $N$ and the
internal dimension $d$ of the surface

\begin{equation}
	L \propto N^{1/d} \ \;.
\end{equation}
This internal dimension $d$ also depends on the external properties of
the surface and $\lambda$ \cite{muenkel92a}.

So let us first look at the specific heat. If the transition is of
second order we would have

\begin{equation}
 C(\lambda ,L) = L^{\alpha / \nu} \hat{C}\left[{ (\lambda -
\lambda_c)L^{-\nu}}\right]
\end{equation}

\noindent where $\hat{C}$ is a scaling function, which depends on how
one implements the surface. At the critical $\lambda$ the scaling function
is regular and the scaling hypothese
leads to
\begin{equation}
	C_N^{max} \propto A N^{\alpha/\nu{d}} + \ldots
\end{equation}

\noindent for the scaling of the peak in the specific heat. If we assume a
first order transition then $C_N^{max}$ diverges as
$L^d$ because of the $\delta$-distribution of
$C_\infty(T)$ \cite{binder86,binder84,landau88}.

An evaluation of the specific heat $C$ of DTRS (neglecting the metric
contribution ${\cal H}_m$) gives the following
expression

\begin{equation}
	C_{all} = \frac{D}{2}  +  \frac{{\lambda}^2}{N} (<{\cal H}_e^2> - <{\cal
H}_e>^2) \ \ \; .
\label{equ:c_fluc}
\end{equation}

\noindent The first part is related to the Gaussian Hamiltonian ${\cal H}_g$
and the second
to the specific heat $C$ of the edge extrinsic curvature ${\cal H}_e$.
The specific heat for the edge extrinsic curvature is shown in
Figure~\ref{fig:ceall} for the two topologies considered.
The interpolation was done
by the method of Ferrenberg and Swendsen \cite{ferrenberg89,ferrenberg88}
using histograms of ${\cal H}_e$ .

\begin{figure}[htbp]
\caption[]
  { \em Specific heat $C$ (edge extrinsic curvature part) of the sphere
        (left part) and of the torus (right part).
  }
\label{fig:ceall}
\end{figure}

\begin{figure}[htbp]
\caption[]
 { \em	Maximum of the specific heat $C_N^{max}$ of the torus $(\Diamond)$
	and the sphere $(\Box)$
 }
\label{fig:cmax_scale}
\end{figure}

Using the data of the specfic heat obtained by applying the extrapolation
method, we can get a very accurate estimate
of the positions of the maxima and of the heights.
Figure~\ref{fig:cmax_scale} shows $C_N^{max}$ of the torus and
the sphere. A change to $L^d$ behaviour
is very unlikely and for that reason the data strongly suggest a continuous
phase transition in agreement with previous work
[8,14-17,38,39].
{}From the data shown in Figure~\ref{fig:cmax_scale} we can obtain the
following upper boundaries of critical exponents

\begin{equation}
	\hbox{Sphere:} \frac{\alpha}{\nu{d}} \leq 0.00 \pm 0.04 \ \ \ \
	\hbox{Torus:} \frac{\alpha}{\nu{d}} \leq 0.06  \pm 0.02
\label{equ:alphnud_res}
\end{equation}
\vspace{1cm}

Using this simple method, we cannot distinguish a diverging specific heat
with very small but positiv $\alpha$, a logarithmic divergence $\alpha_s=0$
and a power law cusp $\alpha_s < 0$, where $\alpha_s$ denotes the exponent of
the singular part of the specific heat. We will use the abbreviation $\alpha$
instead of $\alpha_s$.

Following Fisher \cite{fisher71}, these three cases may be distinguished
with a fit of the form

\begin{equation}
C(\Delta\lambda) = A \cdot \frac{1}{\alpha} \left({{\Delta\lambda}^{-\alpha} -
1}\right) + B
	\ \ \ \; , \ \ \; \Delta\lambda = \left|\lambda_c^N -  \lambda\right|
\label{equ:fit_alpha}
\end{equation}

Figure~\ref{fig:ce_fit_alpha} shows such fits for $\alpha=0.1$ (power law),
$\alpha=-0.01$ (near logarithmic) and $\alpha=-1.2,-2.0$ (power law cusp).
The fits clearly favour a power law cusp
of the specific heat with a large negative value of $\alpha$, although we were
not able to estimate the
exponent $\alpha$ precisely.

\begin{figure}[htbp]
\caption[]
 { \em	Plot of the specific heat $C$ against
	$x = \frac{1}{\alpha} \left({{\Delta\lambda}^{-\alpha} - 1}\right)$ with
	$\alpha=0.1$ (upper left), $-0.01$ (upper right), $-1.2$ (lower left)
	and $\alpha=-2.0$ (lower right).
 }
\label{fig:ce_fit_alpha}
\end{figure}

Our results are in contrast to the estimates of
Renken and Kogut \cite{renken91}
$\alpha/\nu{d}=0.14(3)$ for the sphere. They assumed scaling above
$N=72$, but the more accurate data in Figure~\ref{fig:cmax_scale} shows that
this assumption is not true.
The decreasing slope in Figure~\ref{fig:cmax_scale}
is also present in their Figure 5 \cite{renken91}
although this is slightly masked by
larger error barrs and less data points.
Catteral et all. \cite{catteral91} reported
an estimate of $\alpha/\nu{d}~=~0.185(50)$, but they used a different
discretization based on $\phi^3$ graphs with a flip acceptance rate of only
O(1\%) at $\lambda_c\approx1.5$ and larger correlation times of the edge
extrinsic curvature than in our discretization.

Another possibility to determine the order of the
transition is
the cumulant ${\cal V}_N$ of the edge extrinsic curvature

\begin{equation}
	{\cal V}_N := 1 - \frac{1}{3}
	\frac{{\langle{{{\cal H}_e}^4}\rangle}_N}{{\langle{{{\cal
H}_e}^2}\rangle}_N^2} \;,
\label{equ:VL_def}
\end{equation}

\noindent which behaves quite differently at temperature driven first- and
second-order transitions
\cite{binder86,binder84} :

\begin{eqnarray}
	\hbox{1. and 2.order:} & {\left.{{\cal V}_N}\right|}_{min}
\stackrel{N\rightarrow\infty}{=} \frac{2}{3} & T{\neq}T_c \hbox{fixed}
\label{equ:VL_TneqTc}\\
	\hbox{2.order:} & {\left.{{\cal V}_N}\right|}_{min}
\stackrel{N\rightarrow\infty}{=} \frac{2}{3} & T=T_c(N) \label{equ:VL_TeqTc2}\\
	\hbox{1.order:} & {\left.{{\cal V}_N}\right|}_{min}
\stackrel{N\rightarrow\infty}{=}
		 1-\frac{2\left({E_+^4+E_-^4}\right)}{3{\left({E_+^2+E_-^2}\right)}^2} &
T=T_c(N)\label{equ:VL_TeqTc1}
\end{eqnarray}
$E_+$ and $E_-$ are the energies of the system above and below the
transition. For a very weak first-order transition ($E_+\approx E_-$)
we also have ${\left.{{\cal V}_N}\right|}_{min} \approx 2/3$.

We computed ${\cal V}_N$ defined by equation~(\ref{equ:VL_def}) using again
the method of Ferrenberg and Swendsen \cite{ferrenberg89,ferrenberg88}.
The resulting figures show
the predicted single peak minima with wings
following equation~(\ref{equ:VL_TneqTc}).
The finite size dependence of the
minima of \( {\left.{{\cal V}_N}\right|}_{min} \)
shown in Figure~\ref{fig:VL_min} distinctly favours
the asymptotic behaviour in equation~(\ref{equ:VL_TeqTc2}) and therefore
a continuous phase transition or a very weak first order transition.

\begin{figure}[htbp]
\caption[]
 { \em	Finite size dependence of the minimum
	\( {\left.{{\cal V}_N}\right|}_{min} \) of the reduced cumulant
	of the edge extrinsic curvature. $(\Diamond)$ denotes data of the sphere,
	$(\Box)$ those of the torus and the arrow the large $N$ limit $2/3$ of
	a continuous transition.
 }
\label{fig:VL_min}
\end{figure}

\par In general, finite size scaling predicts also a shift
$\Delta\lambda=\lambda_c^N-\lambda_c^\infty$ of the effective
transition `temperature' $\lambda_c^N$ proportional to $N^{-1}(=L^{-d})$ for a
first- and
proportional to $N^{-1/{\nu}d}(=L^{-1/\nu})$ for a second-order transition.
Unfortunatly $\lambda_c^\infty$ of dynamically triangulated random surfaces is
not known. For that reason,
we have to use a two-parameter fit with unknowns $\lambda_c^\infty$ and
${\nu}d$. But we can improve the reliability of this fit, because we know more.
First, the fit has to be a straight line (neglecting corrections to scaling)
and we have $\lim_{N\rightarrow\infty} \Delta\lambda(N) = 0$.
Second, the shift $\Delta\lambda(N)$ is different for the
specific heat and the cumulant in general. Therefore the slope of the
fitted lines will be different in general, but we still have
$\lim_{N\rightarrow\infty} \Delta\lambda(N) = 0$
for both observables.

\begin{figure}[htbp]
\caption[]
 { \em	Best fit of  $y = \Delta\lambda$ against $x = N^{-1/{\nu}d}$ for
	the sphere (left part) and the torus (right part).
	$(\Diamond)$ denotes the specific heat $C$ data,
	$(\Box)$ those of the cumulant $\cal{V}$.
 }
\label{fig:fit_dlambda}
\end{figure}

Figure~\ref{fig:fit_dlambda} shows the best fits for the sphere and the torus.
The estimates of the parameters are
\begin{equation}
 \lambda_c^\infty = 1.51 \pm 0.04 \ \ \ , \ \ \ {{\nu}d} =  3.2 \pm 0.5
\label{equ:s_alphaandnud_res}
\end{equation}
\noindent for the sphere and
\begin{equation}
 \lambda_c^\infty = 1.47 \pm 0.02 \ \ \ , \ \ \  {{\nu}d} =  2.5 \pm 0.5
\label{equ:t_alphaandnud_res}
\end{equation}
\noindent for the torus.

We turn now to the order parameter itself.
Common practise is to take $\zeta = R/L$ ($R$ is the typical
radius, $L$ the linear size of the membrane) to be a suitable order parameter
[41-47].
Initially \cite{kantor87a} it was also defined as $R_g(L)={\zeta}L$
($L\rightarrow\infty$), with the linear size $L$ of the hexagon and the
radius of gyration
$R_g^2{\propto}\sum_{ij}{\left\langle{{|X_i-X_j|}^2}\right\rangle}$.
A suitable choice for the order parameter therefore is
\begin{equation}
	\zeta := R^2 / N
\end{equation}

\noindent with

\begin{equation}
	R^2 = \frac{1}{N(N - 1)} \left\langle
	\sum_{i,j}^{N}{ \sigma_i \sigma_j {(\vec{X}_i - \vec{X}_j)}^2}
	\right\rangle
\label{equ:r_def_dtrs}
\end{equation}

$\sigma_i$ denotes the number of nearest neighbours, i.e. the number of
links connected to a node $i$.
The associated susceptibility is

\begin{equation}
	\chi_{R^2} := L^d \left( {\langle{\zeta^2}\rangle - {\langle{\zeta}\rangle}^2}
\right)\nonumber\
	= \frac{1}{N} \left( {\langle{R^4}\rangle - {\langle{R^2}\rangle}^2}\right)
\label{equ:chi_def}
\end{equation}

Figure~\ref{fig:s_zeta} shows the order parameter $\zeta$ and
Figure~\ref{fig:s_chi} the susceptibilty of the sphere. The results for the
torus are similar.

\begin{figure}[htbp]
\caption[]
 { \em	Order parameter $\zeta = R^2/N$ of the sphere. Errors are
	smaller than symbol size.
 }
\label{fig:s_zeta}
\end{figure}

\begin{figure}[htbp]
\caption[]
 { \em	Susceptibility $\chi_{R^2}$ of the sphere (Lines to guide the eye).
 }
\label{fig:s_chi}
\end{figure}

\par We also measured another
possible order paramter $\zeta'=\langle{{R'}^2}\rangle / N$,
\[
	\langle{{R'}^2}\rangle = \frac{1}{N} \left\langle
	\sum_{i}^{N}{{\left({\vec{X}_i - \overline{\vec{X}}}\right)}^2}
	\right\rangle
\]
which exhibits a small increase near the phase transition and a slower
decay of $\zeta'$ for $\lambda{\rightarrow}0$.
The difference is caused by a change
of the internal geometry near the phase transition \cite{muenkel92a}.

\par With the data for the sphere in Figure~\ref{fig:s_zeta} and the
corresponding data of the torus one can estimate the critical exponent
$\beta/{\nu}d$ of the order parameter assuming a second order transition.
Here we use as the effective critical temperature the position of the
peak of the specific heat.
$C^{max}$
(c.f. Figure~\ref{fig:rg2_scale}).

\begin{figure}[htbp]
\caption[]
 { \em	Scaling of the radius of gyration squared $\langle{R^2}(N)\rangle$
	at the position of the maximum of the specific heat $C^{max}$.
	$(\Diamond)$ denots the data of the torus, $(\Box)$ data of the sphere.
 }
\label{fig:rg2_scale}
\end{figure}

An estimate of $\beta/{\nu}d$
using

\begin{equation}
	\langle{R^2}(N,\lambda_c^{eff})\rangle \propto N^{\beta/({\nu}d)\ +\ 1}
\end{equation}

\noindent results in

\begin{equation}
	\hbox{Torus:}\ \beta/{\nu}d = 0.28 \pm 0.02\ \;,\; \ \hbox{Sphere:}\
\beta/{\nu}d = 0.35 \pm 0.04
\label{equ:betanud_res}
\end{equation}

\noindent and with the values of ${\nu}d$ in
equation~(\ref{equ:s_alphaandnud_res}) and (\ref{equ:t_alphaandnud_res})

\begin{equation}
	\hbox{Torus:}\ \beta = 0.7 \pm 0.2\ \;,\; \ \hbox{Sphere:}\ \beta = 1.1 \pm
0.2
\label{equ:beta_res}
\end{equation}
\noindent Figure~\ref{fig:chi_scale} shows the scaling of the maxima of the
susceptiblity $\chi$ (equation~(\ref{equ:chi_def}), Figure~\ref{fig:s_chi}).

\begin{figure}[htbp]
\caption[]
 { \em	Scaling of the susceptibility $\chi$
	(equation~(\ref{equ:chi_def}), Fig.~\ref{fig:s_chi}) in the case of
	a torus $(\Diamond)$ and a sphere $(\Box)$.
 }
\label{fig:chi_scale}
\end{figure}
\noindent The results are
\begin{equation}
	\hbox{Torus:}\ \gamma/{\nu}d = 0.62 \pm 0.06\ \;,\; \ \hbox{Sphere:}\
\gamma/{\nu}d = 0.66 \pm 0.06
\label{equ:gammanud_res}
\end{equation}

\noindent and, using the values of ${\nu}d$ in
equation~(\ref{equ:s_alphaandnud_res}) and (\ref{equ:t_alphaandnud_res}),

\begin{equation}
	\hbox{Torus:}\ \gamma = 1.6 \pm 0.5\ \;,\; \ \hbox{Sphere:}\ \gamma = 2.1 \pm
0.5\ .
\label{equ:gamma_res}
\end{equation}

\par With these large error bars and estimated values
$\alpha\approx-1.5$ the results are almost compatible with the scaling relation

\begin{equation}
	\alpha + 2 \beta + \gamma = 2
\label{equ:abg_scaling}
\end{equation}

\vspace{1.0cm}
\noindent{\large\bf Acknowledgements}
\vspace{0.2cm}
\par Partial support by the SFB 123 and the BMFT project $0326657\mbox{D}$ and
$031240284$ is gratefully acknowledged.

\bibliographystyle{aip}

\vfill\eject
\end{document}